\documentclass[conference]{IEEEtran}
\IEEEoverridecommandlockouts
\usepackage{cite}
\usepackage{float} 
\usepackage{graphicx}  
\usepackage{hyperref}

\usepackage{etoolbox}
\usepackage{amsmath,amssymb,amsfonts}
\usepackage{array}
\usepackage{graphicx}
\usepackage{textcomp}
\usepackage{xcolor}
\usepackage{amsmath}
\usepackage{tikz}
\usepackage{nth}
\usepackage{subcaption} 
\usepackage{fancyhdr}
\usepackage{multirow} 
\usepackage{colortbl}
\usepackage{pgf} 

\definecolor{lightgreen}{rgb}{0.85,1.0,0.85}
\definecolor{lightyellow}{rgb}{1.0,1.0,0.6}
\definecolor{lightred}{rgb}{1.0,0.8,0.8}

\newcommand{\colcell}[1]{%
  \pgfmathparse{#1*100}%
  \ifdim \pgfmathresult pt < 50pt \cellcolor{lightred}#1%
  \else
    \ifdim \pgfmathresult pt < 85pt \cellcolor{lightyellow}#1%
    \else \cellcolor{lightgreen}#1%
    \fi
  \fi
}

\usepackage{algorithmicx}
\usepackage{microtype}
\usepackage{siunitx,array,multirow}
\usepackage{booktabs}

\usepackage{tikz}
\usetikzlibrary{shapes.geometric, arrows}

\tikzstyle{startstop} = [rectangle, rounded corners, minimum width=3cm, minimum height=1cm,text centered, draw=black, fill=gray!20]
\tikzstyle{process} = [rectangle, minimum width=3.5cm, minimum height=1cm, text centered, draw=black, fill=blue!10]
\tikzstyle{arrow} = [thick,->,>=stealth]

\usepackage{tabularx}
\usepackage{multirow}
\usepackage{threeparttable}
\usepackage{multicol}
\usepackage{amsmath}
\usepackage{makecell} 

\usepackage[english]{babel}
\usepackage[utf8]{inputenc}
\usepackage{algorithm}
\usepackage[noend]{algpseudocode}

\newcommand{\subsubsubsection}[1]{\paragraph{#1}\mbox{}\\}
\usetikzlibrary{shapes,arrows}
\usepackage{verbatim}

\usepackage{url}
\usepackage{amsmath}
\usepackage{textcomp}
\usepackage{siunitx}
\usepackage[utf8]{inputenc}
\usepackage{upgreek}

\makeatletter
 \let\old@ps@headings\ps@headings
 \let\old@ps@IEEEtitlepagestyle\ps@IEEEtitlepagestyle
 \def\confheader#1{%

 \def\ps@IEEEtitlepagestyle{%
 \old@ps@IEEEtitlepagestyle%
 \def\@oddhead{\strut#1\hfill\strut}%
 \def\@evenhead{\strut\hfill#1\hfill\strut}%
 }%
 \ps@headings%
 }
 \makeatother

 \confheader{
    \noindent  
    \fontsize{9}{11}\selectfont  
    \parbox{\textwidth}{ 
        2025 International Conference on Quantum Photonics, Artificial Intelligence, and Networking (QPAIN) \\ 
        31 July - 2 August 2025, Rangpur, Bangladesh
    }
}

\begin{document}
\title{Handling Class Imbalance Problem in Skin Lesion Classification: Finding Strengths and Weaknesses of Various Balancing Techniques} 
\author {\IEEEauthorblockN{Ariful Islam Khandaker\IEEEauthorrefmark{1}$^{1}$, Abdullah Al Shafi\IEEEauthorrefmark{2}$^{1}$, and Mohiuddin Ahmad\IEEEauthorrefmark{3}}
\IEEEauthorblockA{
Institute of Information and Communication Technology, Khulna University of Engineering \& Technology, Khulna, Bangladesh\IEEEauthorrefmark{1}\IEEEauthorrefmark{2}\\
Department of Computer Science \& Engineering, Khulna University of Engineering \& Technology, Khulna, Bangladesh\IEEEauthorrefmark{2}
\\Department of Electrical and Electronic Engineering, Khulna University of Engineering \& Technology, Khulna, Bangladesh\IEEEauthorrefmark{3}\\
aikhandaker@iict.kuet.ac.bd\IEEEauthorrefmark{1}, abdullah@iict.kuet.ac.bd\IEEEauthorrefmark{2}}, ahmad@eee.kuet.ac.bd\IEEEauthorrefmark{3}
\thanks{$^{1}$Equal contribution}}

\maketitle
\begin{abstract}
Automatic skin lesion classification from dermoscopy images is important for the early diagnosis of skin diseases such as melanoma. Class imbalance in skin lesion datasets, notably the defects in the representation of malignant(cancerous) cases, is one of the difficulties for deep learning models' performances and generalizations. This paper offers an exhaustive review of some of the balancing methods that aim to address class imbalances using the example of the ISIC 2016 dataset. A light-weight CNN model, MobileNetV2, was combined with under-sampling, over-sampling, and hybrid balancing methods such as Tomek Links(TL), SMOTE, and SMOTE with TL. Over-sampling methods like SMOTE and ADASYN improve performance but may lead to overfitting due to redundant synthetic samples. Hybrid methods like SMOTE+TL counter this drawback by removing noisy or boundary samples so that model generalization is enhanced. Thus, this analysis stresses the need to choose the right balancing methods for robust and sensitive diagnostic systems in medical image processing.
\end{abstract}


\section{Introduction}
The classification of skin lesions is a very critical factor that allows early detection and diagnosis of a variety of dermatological disorders including skin cancer\cite{rastgoo2016tackling}. The increasing number of skin-related disorders, in particular melanomas, has, however, raised the need for precise and automated diagnostic tools\cite{yao2021single}. These are ones that expert dermatologists would traditionally make manual diagnostic tests, and they tend to be very laborious and subjective\cite{kaur2022melanoma}. But prior identification of diseases has significant implications for managing the outcomes of treatment, affordability of healthcare costs, and quality of life of patients suffering from the condition\cite{zannat2025bridging}. In recent years, computer vision has offered great promise in classifying skin lesions using dermoscopic images in real time\cite{yao2021single}.

Convolutional Neural Networks (CNNs) are superior to many manually created feature-based techniques in learning discriminative characteristics automatically\cite{yao2021single}. However, imbalanced datasets with differences between classes that are small and large variations within classes still make classification performance miserable\cite{rastgoo2016tackling}.
Dermoscopy image datasets commonly suffer from severe class imbalance, with benign (noncancerous) instances grossly outnumbering malignant (cancerous) cases. Such a disproportionate distribution is a fundamental challenge for supervised algorithms as the models tend to become biased to the dominance classes during training. So, the minority classes, which are usually the most important ones such as melanoma, are under-predicted, resulting in a decrease of sensitivity and generalization capacity\cite{yao2021single}.

Using simple accuracy as a performance metric becomes unreliable under these conditions, thereby inspiring researchers to use more informative metrics such as precision, recall, and F1-score\cite{owusu2023imbalanced}. Accordingly, many ways exist to balance the proportions, including data augmentation\cite{yao2021single}, under-sampling\cite{rastgoo2016tackling}, over-sampling\cite{rastgoo2016tackling}, and feature selection methods\cite{edward2024comprehensive}. However, these procedures come with their own limitations, for example: the chances that the model might get overfitted and that generated samples might be unrealistic, especially when the minority class is very scarce. Therefore, tackling the problems of imbalanced data is necessary to create an effective and trustworthy clinial deep learning model for dermoscopic image analysis.

The intense focus on appropriately tackling class imbalance in skin lesion classification encompasses enhancement of model performance by compensating for the lack of availability of malignant(cancerous) samples. Rastgoo et al.\cite{rastgoo2016tackling} conducted an extensive study on data balancing techniques for skin lesion classification, showing the effectiveness of under-sampling methods such as NearMiss-2. The method provided in \cite{yao2021single} works as a single DCNN that utilizes RandAugment, MWNL, and cumulative learning to get superior performances against ensembles on small, imbalanced datasets. While presenting their recent research \cite{owusu2023imbalanced}, the authors suggest a new evaluation method PMEA, which combines TPR with TNR, to offset disadvantages of using prediction accuracy to judge classifiers built from imbalanced data, especially in health contexts.  In \cite{edward2024comprehensive}, a multi-class rebalancing framework using domain-specific medical tests such as SCUT, SHAP-RFE feature selection, and DES-MI was put forward for improving performance on medical datasets.\\
The following are the contributions that our work has made:

(i) We present a comparative study of a number of balancing techniques applied for the classification of skin lesions. These include oversampling, undersampling, hybrid techniques, and ensemble learning.

(ii) It studies the effect of these balancing methods on the classifier performance of the CNNs on imbalanced data of medical images.

(iii) The study presents a systematic evaluation to find the strengths and weaknesses of these balance techniques in medical image classification.

(iv) It uses a real-world skin lesion dataset to study the effect of data imbalance on model performance and explain why some methods help the detection of the minority classes at the expense of increased noise or overfitting.

(v) Evaluation of balancing methods is conducted in the study based on precision, recall, and F1-score, thus providing a fairly balanced view of both overall and class-wise detection performance.

(vi) It offers guidance on selecting suitable balancing techniques based on dataset characteristics and model behavior based on their pros and cons.

This paper's remaining sections are structured in the following manner: In Sect. \ref{sec:dataset}, the dataset (ISIC 2016) is described with an imbalanced class distribution. Our proposed system architecture is described in Sect. \ref{sec:methodology}. The classification system intended to study data balancing strategies is summarized in Sect.\ref{sec:balancing} and in Sect. \ref{sec:result}, the validation and quantitative assessment are covered, followed by a conclusion(sect. \ref{sec:conclusion}).

\section{Dataset}
\label{sec:dataset}
\begin{figure}[htbp]
\centerline{\includegraphics[scale=.40]{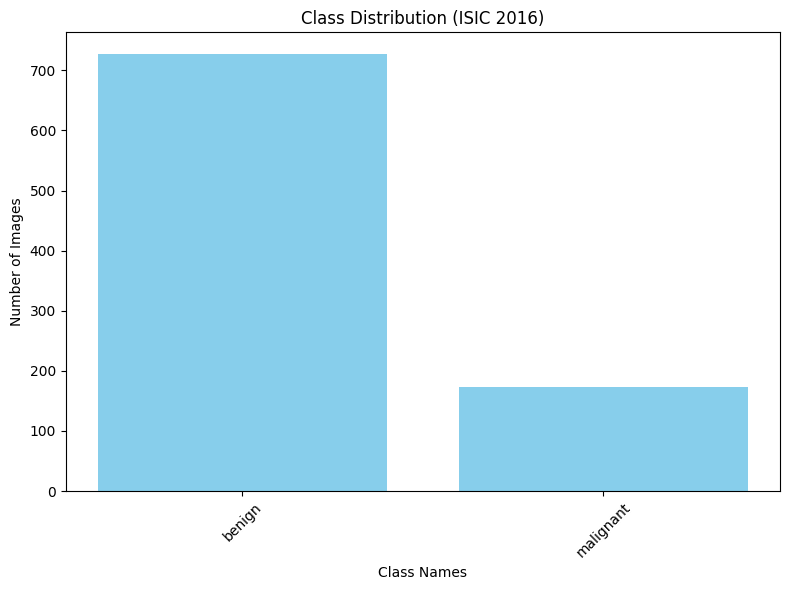}}
\caption{Data distribution of ISIC 2016 dataset. The distribution shows that the dataset is incredibly unbalanced. The amount of benign(non-cancerous) images is much higher than that of malignant(cancerous) images.}
\label{fig:dd}
\end{figure}

We have used ISIC 2016\cite{gutman2016skin} dataset to carry out our research. Fig. \ref{fig:dd} shows the data imbalance problem present in this dataset.

\section{Proposed Methodology}
\label{sec:methodology}
\subsection{Balancing}
Various balancing techniques in \ref{sec:balancing} were applied to the dataset and compared to find the optimum one.

\subsection{Resizing}
For the sake of consistency in the input size as well as in improving training stability of models, images get resized first before inputting to the network. In this work, all the images are resized into a fixed uniform size of 224×224 pixels using a resizing layer. It is necessary since most CNNs look for the standard-sized input.

\subsection{Rescaling}
Once resized, the pixel intensities of the images, are adjusted into the range of [0, 1] after initially being in the range [0, 255] using a rescaling factor of 1/255. 
Normalization helps to ensure better convergence while training and numerical stability within the network.



\subsection{Augmentation}
To increase the diversity of the training data and boost model generalization, data augmentation approaches are used. 
These improvements simulate real-world distortions and variability, making the model more robust to new, unseen data.  





\subsection{Classification Model}
The classification is performed with a lightweight convolutional neural network (CNN) based on MobileNetV2. MobileNetV2\cite{sandler2018mobilenetv2} is a lightweight CNN architecture for on-device vision tasks. It is based on depthwise separable convolutions, a technique that reduces the computational cost without compromising much accuracy. MobileNetV2 model uses an inverted residual with a linear bottleneck that improves performance and efficiency. The architecture is particularly suitable for edge and mobile devices. MobileNetV2 is usually pre-trained on a large dataset like ImageNet for classification tasks and fine-tuned on specific tasks to deploy quickly in edge environments.

\begin{algorithm}
\caption{Classification using MobileNetV2}
\begin{algorithmic}[1]
\State \textbf{Input:} Raw input images (Dataset)
\State Apply various balancing techniques to handle data imbalance problem
\State Resize each input image to $224 \times 224$ pixels
\State Adjust pixel values to fall inside the range $[0,1]$ (Normalization)
\State Load MobileNetV2 pretrained on ImageNet with \texttt{include\_top=False}
\State Freeze all layers of the base MobileNetV2 model
\State Pass input images through MobileNetV2 to extract features
\State Apply Global Average Pooling to reduce feature maps
\State Incorporate a 128-unit dense layer with ReLU activation.
\State For binary classification, insert a final dense layer with one unit and sigmoid activation.
\State Build the model using "BinaryCrossentropy" as the loss function and "Adam" as the optimizer.
\State Use batch size 32 and train the model for 40 epochs.
\State \textbf{Output:} Trained model for binary classification
\end{algorithmic}
\end{algorithm}

\section{Balancing Strategies}
\label{sec:balancing}
Equalizing both majority and minority class samples is the task of data balancing.

\subsection{Feature Space Sampling}
The issue of an imbalanced dataset can be resolved in three ways: (1) US, (2) OS, and (3) a combination of the two.

\subsubsection{Under-Sampling(US)}
In US, to equal the amount of samples from the minority class, the majority class's sample size is decreased. The following techniques are used to achieve this balancing:

\subsubsubsection{Random Under-Sampling (RUS)}
Random selection without replacement is used to choose a subset of samples from the majority class to achieve RUS\cite{rastgoo2016tackling}, which leaves an equal number of majority and minority class samples.


\subsubsubsection{Tomel Link(TL)}
The majority class of the original dataset can be under-sampled using TL \cite{tomek1976two} (Algorithm \ref{algo:tl}).



\vspace{-0.4em}
\begin{algorithm}[H]
\caption{TL Under-Sampling}
\label{algo:tl}
\textbf{Input:} $D_{\text{maj}}$ -- list of majority class samples, $D_{\text{min}}$ -- list of minority class samples, $k$ -- \# of NN(nearest neighbors)\\
\textbf{Output:} $D_{\text{re}}$ -- under-sampled dataset
\begin{algorithmic}[1]
\State $D_{\text{re}} \gets D_{\text{maj}} \cup D_{\text{min}}$
\For{each sample $x_i$ in $D_{\text{maj}}$}
    \State $x_j \gets$ Nearest\_neighbor($x_i$)  
    \If{Class($x_i$) $\neq$ Class($x_j$)} 
        \If{No closer neighbor for $x_i$ in $D_{\text{maj}}$ and no closer neighbor for $x_j$ in $D_{\text{min}}$}
            \State Remove $x_i$ from $D_{\text{re}}$  
        \EndIf
    \EndIf
\EndFor
\State \Return $D_{\text{re}}$ 
\end{algorithmic}
\end{algorithm}
\vspace{-1.0em}

\subsubsubsection{Near Miss(NM)}
\vspace{-0.3em}
According to Mani and Zhang \cite{mani2003knn}, NearMiss(NM) provides three distinct techniques for undersampling the majority class: NM1, NM2, and NM3. NM1 chooses majority class samples in order to minimize the average distance between each sample and the k NN samples from the minority class. NM2, conversely, preserves the majority samples that are far from the minority samples. NM3 can thus be thought of as a compromise between NM1 and NM2 but with some added focus. First, NM3 identifies a predetermined number of samples of the majority class that are most similar to each sample of the minority class. From this selection, only those majority samples that are farthest from the minority class (on average) are retained.

\subsubsubsection{Neighborhood Cleaning Rule(NCR)}
\vspace{-0.2em}
NCR\cite{rastgoo2016tackling} discards misleading or noisy samples to improve the data quality. It verifies k-nearest neighbors(k=3) of an instance.

- The majority sample is removed when it is different from its neighbors.

- Minority sample gets rid of its neighbors (majority in most cases) if it falls outside the cluster.

    

\subsubsubsection{Clustering Under Sampling(CUS)}

\vspace{-0.8em}

The CUS\cite{yan2022spatial} algorithm selects the centroids of the clusters formed by grouping the majority samples into clusters equal to the amount of minority samples using k-means clustering. To create a balanced dataset, the centroids of all minority samples and majority clusters are eventually combined.

\subsubsection{Over-Sampling(OS)}
To balance the amount of samples in both groups, OS is carried out by creating new samples from the minority class.

\subsubsubsection{Random Over-sampling(ROS)}

\vspace{-0.8em}
To balance the dataset, samples from the minority class are duplicated using the ROS procedure. Until both groups reach the same size, replacements are used to select random samples from the minority class\cite{rastgoo2016tackling}. 

\subsubsubsection{SMOTE}
SMOTE\cite{chawla2002smote} is an approach for creating synthetic samples in the feature space (Algorithm \ref{algo:smote}). We set k to 3 in our study.



\begin{algorithm}
\caption{SMOTE}
\label{algo:smote}
\textbf{Input:} $D_{\text{maj}}$ -- list of majority class samples, $D_{\text{min}}$ -- list of minority class samples, $k$ -- \# of NN(nearest neighbors)\\
\textbf{Output:} $D_{\text{re}}$ -- over-sampled dataset

\begin{algorithmic}[1]
\ForAll{$x_i \in D_{\text{min}}$}
    \State $NN_k \gets$ \Call{KNNs}{$x_i, D_{\text{min}}, k$}
    \State $x_{nn} \gets$ \Call{RandomSample}{$NN_k$}
    \State $\sigma \gets$ \Call{RandomNumber}{$[0, 1]$}
    \State $x_j \gets x_i + \sigma \cdot (x_{nn} - x_i)$
    \State \Call{Add}{$x_j, D_{\text{os}}$}
\EndFor
\State $D_{\text{re}} \gets D_{\text{os}} \cup D_{\text{min}} \cup D_{\text{maj}}$
\State \Return $D_{\text{re}}$
\end{algorithmic}
\end{algorithm}

\begin{table*}[!ht]
\centering
\setlength\extrarowheight{5pt}
\caption{Classification results on the test set of ISIC 2016. Color coding (lightgreen: High ($\ge$ 0.85), lightyellow: Moderate (0.50–0.84), lightred: Low ($<$ 0.50) ) is used for better readability. The balancing techniques are colored based on macro precision, recall and f1-score.}
\label{tab:result}
\resizebox{\textwidth}{!}{%
\begin{tabular}{|c|c|c|c|c|c|c|c|c|c|c|c|c|c|}
\hline
\multirow{2}{*}{\textbf{Balancing techniques}} &
  \multirow{2}{*}{\textbf{Accuracy}} &
  \multicolumn{4}{c|}{\textbf{Precision}} &
  \multicolumn{4}{c|}{\textbf{Recall}} &
  \multicolumn{4}{c|}{\textbf{F1-score}} \\ \cline{3-14} 
& & \textbf{Benign} & \textbf{Malignant} & \textbf{Macro} & \textbf{Micro} 
  & \textbf{Benign} & \textbf{Malignant} & \textbf{Macro} & \textbf{Micro} 
  & \textbf{Benign} & \textbf{Malignant} & \textbf{Macro} & \textbf{Micro} \\ \hline
\rowcolor{lightred}
Imbalanced (IB) & \colcell{0.81} & \colcell{0.81} & \colcell{0.00} & \colcell{0.41} & \colcell{0.81} & \colcell{1.00} & \colcell{0.00} & \colcell{0.50} & \colcell{0.81} & \colcell{0.90} & \colcell{0.00} & \colcell{0.45} & \colcell{0.81} \\ \hline

\rowcolor{lightyellow}
RUS & \colcell{0.69} & \colcell{0.77} & \colcell{0.62} & \colcell{0.69} & \colcell{0.69} & \colcell{0.64} & \colcell{0.75} & \colcell{0.69} & \colcell{0.69} & \colcell{0.70} & \colcell{0.68} & \colcell{0.69} & \colcell{0.69} \\

\rowcolor{lightgreen}
TL & \colcell{0.93} & \colcell{0.92} & \colcell{1.00} & \colcell{0.96} & \colcell{0.93} & \colcell{1.00} & \colcell{0.57} & \colcell{0.79} & \colcell{0.93} & \colcell{0.96} & \colcell{0.73} & \colcell{0.85} & \colcell{0.93} \\

\rowcolor{lightyellow}
NM1 & \colcell{0.64} & \colcell{0.64} & \colcell{0.64} & \colcell{0.64} & \colcell{0.64} & \colcell{0.58} & \colcell{0.70} & \colcell{0.64} & \colcell{0.64} & \colcell{0.61} & \colcell{0.67} & \colcell{0.64} & \colcell{0.64} \\
\rowcolor{lightyellow}
NM2 & \colcell{0.67} & \colcell{0.62} & \colcell{0.69} & \colcell{0.65} & \colcell{0.67} & \colcell{0.33} & \colcell{0.88} & \colcell{0.60} & \colcell{0.67} & \colcell{0.43} & \colcell{0.77} & \colcell{0.60} & \colcell{0.67} \\

\rowcolor{lightyellow}
NM3 & \colcell{0.77} & \colcell{0.63} & \colcell{0.93} & \colcell{0.78} & \colcell{0.77} & \colcell{0.92} & \colcell{0.68} & \colcell{0.80} & \colcell{0.77} & \colcell{0.75} & \colcell{0.78} & \colcell{0.76} & \colcell{0.77} \\
\rowcolor{lightyellow}
CUS & \colcell{0.66} & \colcell{0.94} & \colcell{0.55} & \colcell{0.74} & \colcell{0.66} & \colcell{0.44} & \colcell{0.96} & \colcell{0.70} & \colcell{0.66} & \colcell{0.60} & \colcell{0.70} & \colcell{0.65} & \colcell{0.66} \\
\rowcolor{lightred}
NCR & \colcell{0.63} & \colcell{0.64} & \colcell{0.54} & \colcell{0.59} & \colcell{0.63} & \colcell{0.93} & \colcell{0.15} & \colcell{0.54} & \colcell{0.63} & \colcell{0.76} & \colcell{0.23} & \colcell{0.49} & \colcell{0.63} \\ \hline
\rowcolor{lightyellow}
ROS & \colcell{0.72} & \colcell{0.72} & \colcell{0.72} & \colcell{0.72} & \colcell{0.72} & \colcell{0.80} & \colcell{0.63} & \colcell{0.72} & \colcell{0.72} & \colcell{0.76} & \colcell{0.67} & \colcell{0.72} & \colcell{0.72} \\

\rowcolor{lightgreen}
SMOTE & \colcell{0.88} & \colcell{0.80} & \colcell{1.00} & \colcell{0.90} & \colcell{0.88} & \colcell{1.00} & \colcell{0.74} & \colcell{0.87} & \colcell{0.88} & \colcell{0.89} & \colcell{0.85} & \colcell{0.87} & \colcell{0.88} \\

\rowcolor{lightgreen}
ADASYN & \colcell{0.88} & \colcell{0.82} & \colcell{0.99} & \colcell{0.90} & \colcell{0.88} & \colcell{0.99} & \colcell{0.75} & \colcell{0.87} & \colcell{0.88} & \colcell{0.90} & \colcell{0.85} & \colcell{0.88} & \colcell{0.88} \\ \hline

\rowcolor{lightgreen}
SMOTE+TL & \colcell{0.86} & \colcell{0.77} & \colcell{1.00} & \colcell{0.89} & \colcell{0.86} & \colcell{1.00} & \colcell{0.72} & \colcell{0.86} & \colcell{0.86} & \colcell{0.87} & \colcell{0.84} & \colcell{0.87} & \colcell{0.86} \\

\rowcolor{lightgreen}
SMOTE+ENN & \colcell{0.85} & \colcell{0.82} & \colcell{1.00} & \colcell{0.91} & \colcell{0.85} & \colcell{1.00} & \colcell{0.70} & \colcell{0.82} & \colcell{0.85} & \colcell{0.90} & \colcell{0.82} & \colcell{0.87} & \colcell{0.85} \\ \hline
\rowcolor{lightred}
Ensemble (Bagging) & \colcell{0.66} & \colcell{0.85} & \colcell{0.14} & \colcell{0.50} & \colcell{0.66} & \colcell{0.72} & \colcell{0.26} & \colcell{0.49} & \colcell{0.66} & \colcell{0.78} & \colcell{0.19} & \colcell{0.48} & \colcell{0.66} \\ \hline

\end{tabular}%
}
\end{table*}

\subsubsubsection{ADASYN}
The ADASYN\cite{he2008adasyn} algorithm works basically on generating new synthetic samples for the minority class according to the classification difficulty. For this purpose, the classification difficulty, or imbalance degree, is calculated for each minority sample relative to its neighbors. More synthetic samples are created for those minority samples that are more difficult to classify.

\subsubsection{Combination of OS and US}
OS techniques can be coupled with US techniques to reduce the problem of overfitting.
\subsubsubsection{SMOTE+TL}
Removing the TL from the majority and minority classes can prevent overfitting caused by SMOTE oversampling \cite{rastgoo2016tackling}.



\subsubsubsection{SMOTE+ENN}
For the same reason as to prevent overfitting, SMOTE and Edited Nearest Neighbor(ENN) are merged\cite{rastgoo2016tackling}.

\subsection{Ensemble Learning}
Bagging or bootstrap aggregating\cite{galar2011review} entails building several models on various balanced subsets of the training data and then combining their predictions. A base model is trained on each subset, and finally the prediction is made by majority voting across all models.



\section{Experimental Analysis and Results}
\label{sec:result}

\subsection{Experimental Setup}
With Pandas, NumPy, Matplotlib, TensorFlow, and the CUDA Toolkit for the acceleration of computations on the GPU, the presented work was conducted in Python. Three divisions of the datasets were created: 80\% training, 10\% validation, and 10\% testing.

\begin{table*}[!ht]
\centering
\setlength\extrarowheight{6pt}
\caption{Comparison of the strengths and weaknesses of various balancing techniques for ISIC 2016 dataset. This table complements color coding of Table \ref{tab:result}, and both can be employed alongside for a complete perspective.}
\label{tab:comparison}
\resizebox{\textwidth}{!}{
\begin{tabular}{|c|c|c|}
\hline
\textbf{Method} & \textbf{Strengths} & \textbf{Weaknesses} \\ \hline
\cellcolor{lightred}Imbalanced & Simple and fast; no modification to the data & Very skewed, biased predictions \\ \hline
\cellcolor{lightyellow}RUS & Removes ungainly samples & Loss of relevant sample space can lead to under-training \\ 
\cellcolor{lightgreen}TL & Ensures clearer discrimination between the two classes & Might incur a lot of data loss \\
\cellcolor{lightyellow}NM1 & Eliminates noisy samples & Leads to the risk of underfitting \\
\cellcolor{lightyellow}NM2 & Efficiently manages noisy and overlapping samples & May remove too many samples \\
\cellcolor{lightyellow}NM3 & Improves model accuracy by cleaning the noisy data & High computational cost \\
\cellcolor{lightyellow}CUS & Maintains dataset structure while balancing classes & Could result in underfitting \\
\cellcolor{lightred}NCR & Effective with intricate datasets & Performance is significantly impacted by parameter adjustments \\ \hline
\cellcolor{lightyellow}ROS & Increases minority class samples without sacrificing information & Can result in overfitting \\
\cellcolor{lightgreen}SMOTE & Generates synthetic samples while preserving the diversity of data & Overfitting risk, particularly for small datasets \\
\cellcolor{lightgreen}ADASYN & Improves SMOTE by focusing on classification difficulty samples & Costly to compute and could result in overfitting \\ \hline
\cellcolor{lightgreen}SMOTE+TL & Reduces overfitting more than SMOTE-only & Can remove valuable minority class samples \\
\cellcolor{lightgreen}SMOTE+ENN & Improved decision boundaries; Prevents overfitting & May result in loss of data and underfitting \\ \hline
\cellcolor{lightred}Ensemble (Bagging) & Combines multiple models for better performance & High computational cost and complexity in implementation \\ \hline
\end{tabular}
}
\end{table*}

\subsection{Evaluation Metrics}
Accuracy might be deceptive when used to imbalanced datasets. Precision, recall, and other valuable metrics, such as the F1 score, come into play under these circumstances. 

\subsection{Result Analysis}

\begin{figure}[H]
\centerline{\includegraphics[scale=.5]{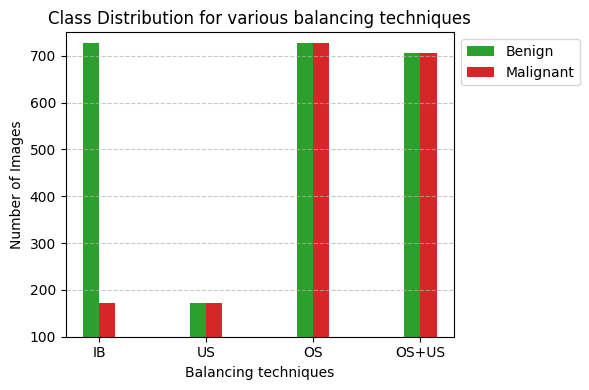}}
\caption{Data distribution of various balancing techniques.}
\label{fig: class_distribution_balance}
\end{figure}

The distribution of classes for various balancing categotiries are shown in Fig. \ref{fig: class_distribution_balance}. However, different balancing techniques demonstrate varying impacts on overall performance (Table \ref{tab:result}). Training on the imbalanced dataset gives high overall accuracy but fails to identify malignant(minority) cases completely.


RUS provides better class balance but reduces overall accuracy by losing information. TL possesses high precision and accuracy for malign instances, but recall is not high. NearMiss approaches (NM1, NM2, NM3), particularly NM3 and CUS show average performance. NCR is not sufficiently encouraging minority class identification.

ROS balances classes moderately but overfits. On the other hand, SMOTE and ADASYN achieve the best overall performance, with high F1-scores, precision, and recall, and thus are appropriate for safety-critical applications.

\begin{figure}[H]
\centerline{\includegraphics[scale=.6]{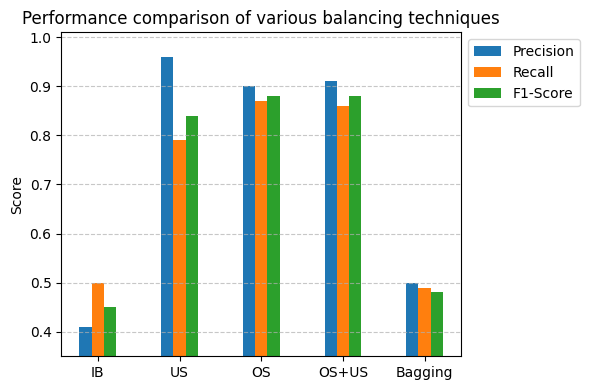}}
\caption{Performance comparison of US, OS and their combination. Here, we take the best techniques of each category. OS + US results from a trade-off between OS and US.}
\label{fig: performance_comp}
\end{figure}

\begin{figure}[H]
     \centering
     \begin{subfigure}[b]{0.33\textwidth}
         \centering
         \includegraphics[width=\textwidth]{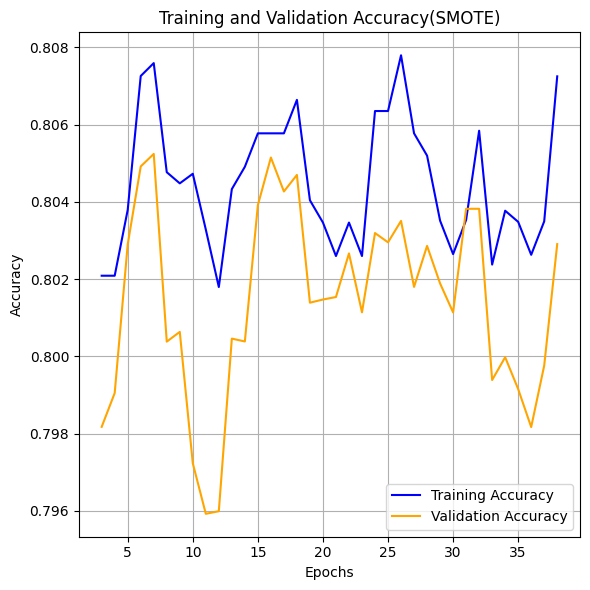}
         \caption{}
         \label{smote}
     \end{subfigure}
     \begin{subfigure}[b]{0.33\textwidth}
         \centering
         \includegraphics[width=\textwidth]{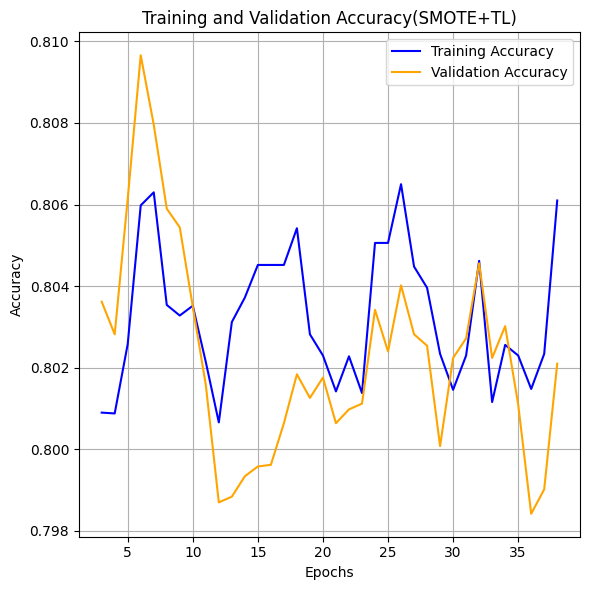}
         \caption{}
         \label{smote+tl}
     \end{subfigure}
        \caption{Comparison of Training and Validation Accuracy: (a) SMOTE-only vs (b) SMOTE+TL. SMOTE+TL stabilizes and prevents overfitting, resulting in improved performance on new data.
}
\label{learing curve}
\end{figure}

Hybrid approaches like SMOTE+TL and SMOTE+ENN perform comparable with pure SMOTE or ADASYN but maintain adequate balance and prevent overfitting. Fig. \ref{fig: performance_comp} demonstrates the trade-off between OS and US achieved by these approaches.

Finally, Ensemble (Bagging) is poor with low malignant (minority) class detection, highlighting the challenge of ensemble methods in highly imbalanced medical data.


Training and validation accuracies for both methods, SMOTE-only and SMOTE+TL, are displayed in Fig. \ref{learing curve}. In (a), the performance of the SMOTE-only model exhibits wild oscillations in validation accuracies, which may indicate overfitting. The latter (b) does, however, show better alignment between the two curves, supporting the notion that combining TL improves harmony and generalization. Nonetheless, the SMOTE+TL setting appears more stable and consistent throughout the epochs.

In the experiment, it is evident that each balancing strategy required varying computation demands. SMOTE, ADASYN, and hybrids like SMOTE+ENN require relatively long execution times, due to synthetic sample generation and nearest-neighbor computations. ROS and RUS had a comparatively light overhead.

Table \ref{tab:comparison} summarizes the overall findings of our experiment. In summary, SMOTE and ADASYN show the highest performance in classification with high precision, recall, and F1 scores but may overfit. Other methods like RUS and NearMiss improve balance but at the cost of lower accuracy. Although the positive side of hybrid techniques such as SMOTE+ENN include better boundary refinement and overfitting mitigation, they usually have a downside of being more compute-intensive and requiring significant hyperparameter tuning, unlike simpler methods such as SMOTE or ADASYN. Such a trade-off between better generalization at the cost of computation has to be given due thought when finally deploying these algorithms, especially for resource-limited settings.

\begin{figure}[htbp]
\centerline{\includegraphics[scale=.50]{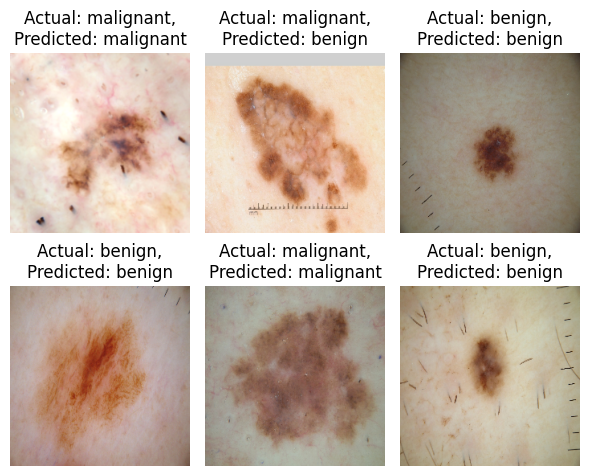}}
\caption{Sample prediction on ISIC 2016 dataset using SMOTE+ENN as Balancing techniques.}
\label{fig: prediction}
\end{figure}

Fig. \ref{fig: prediction} shows prediction on a set of sample images where we find out that 5 are predicted correctly out of 6, which is a good indication. The same predictions were found for the dominant balancing techniques (according to performance) like TL, SMOTE, ADASYN, and TL+SMOTE. 

Table \ref{comparison} shows a comparison between the present work and some recent relevant studies. It shows that our study outperformed other recent works on skin lesion classification utilizing ISIC 2016 dataset.

\begin{table}[htbp]
    \centering
    \caption{A comparison of the classification results with several recent pertinent studies}
    \label{comparison}
    \begin{tabular}{|c|c|c|c|c|}
        \hline
        \textbf{Authors} & \textbf{PRE} & \textbf{REC} & \textbf{F1-score} \\
        \hline
        Kaur et al. \cite{kaur2022melanoma} & 0.82 & 0.81 & 0.82\\
        \hline
        Gun et al.\cite{10757164} & 0.82 & 0.83 & 0.80\\
        \hline
        Al Shafi et al.\cite{al2025skin} & 0.88 & 0.84 & 0.86\\
        \hline
        \textbf{Present study} & \textbf{0.90} & \textbf{0.87} & \textbf{0.88}\\
        \hline
    \end{tabular}
\end{table}


\section{Conclusions}
\label{sec:conclusion}
This research thoroughly examines the impact of class imbalance on the performance of skin lesion classification models based on deep learning. Our experiments with various balancing techniques on the ISIC 2016 dataset showed that conventional methods like RUS or ROS may perform poorly due to information loss or overfitting. On the other hand, advanced techniques like SMOTE and ADASYN yield much better classification metrics, making them more suitable for critical medical applications, where being sensitive to minority classes is crucial while running the risk of overfitting. Moreover, combining oversampling with undersampling techniques (e.g., SMOTE+TL) establishes an appropriate compromise between generalization and overfitting. Practitioners may find these insights useful when deciding on the type of balancing they want to employ in building a stable skin lesion classifier for clinical use. Although our work is centered around binary classification, the results obtained from the analysis of balancing methods should be applicable in multi-class and multi-label classification problems that are common in medical image analysis. However, because of resource constraint, we were not able to apply the all the balancing techniques on larger dataset like HAM10000 or ISIC 2020. In the future, we would like to do so. We also like to explore more such balancing techniques.




\bibliographystyle{./IEEEtran}

\begin{thebibliography}{10}
\providecommand{\url}[1]{#1}
\csname url@samestyle\endcsname
\providecommand{\newblock}{\relax}
\providecommand{\bibinfo}[2]{#2}
\providecommand{\BIBentrySTDinterwordspacing}{\spaceskip=0pt\relax}
\providecommand{\BIBentryALTinterwordstretchfactor}{4}
\providecommand{\BIBentryALTinterwordspacing}{\spaceskip=\fontdimen2\font plus
\BIBentryALTinterwordstretchfactor\fontdimen3\font minus \fontdimen4\font\relax}
\providecommand{\BIBforeignlanguage}[2]{{%
\expandafter\ifx\csname l@#1\endcsname\relax
\typeout{** WARNING: IEEEtran.bst: No hyphenation pattern has been}%
\typeout{** loaded for the language `#1'. Using the pattern for}%
\typeout{** the default language instead.}%
\else
\language=\csname l@#1\endcsname
\fi
#2}}
\providecommand{\BIBdecl}{\relax}
\BIBdecl

\bibitem{rastgoo2016tackling}
M.~Rastgoo, G.~Lemaitre, J.~Massich, O.~Morel, F.~Marzani, R.~Garcia, and F.~Meriaudeau, ``Tackling the problem of data imbalancing for melanoma classification,'' in \emph{Bioimaging}, 2016.

\bibitem{yao2021single}
P.~Yao, S.~Shen, M.~Xu, P.~Liu, F.~Zhang, J.~Xing, P.~Shao, B.~Kaffenberger, and R.~X. Xu, ``Single model deep learning on imbalanced small datasets for skin lesion classification,'' \emph{IEEE transactions on medical imaging}, vol.~41, no.~5, pp. 1242--1254, 2021.

\bibitem{kaur2022melanoma}
R.~Kaur, H.~GholamHosseini, R.~Sinha, and M.~Lind{\'e}n, ``Melanoma classification using a novel deep convolutional neural network with dermoscopic images,'' \emph{Sensors}, vol.~22, no.~3, p. 1134, 2022.

\bibitem{zannat2025bridging}
R.~Zannat, A.~Al~Shafi, and A.~Muntakim, ``Bridging the gap in bangla healthcare: Machine learning based disease prediction using a symptoms-disease dataset,'' in \emph{2025 International Conference on Electrical, Computer and Communication Engineering (ECCE)}.\hskip 1em plus 0.5em minus 0.4em\relax IEEE, 2025, pp. 1--6.

\bibitem{owusu2023imbalanced}
M.~Owusu-Adjei, J.~Ben Hayfron-Acquah, T.~Frimpong, and G.~Abdul-Salaam, ``Imbalanced class distribution and performance evaluation metrics: A systematic review of prediction accuracy for determining model performance in healthcare systems,'' \emph{PLOS Digital Health}, vol.~2, no.~11, p. e0000290, 2023.

\bibitem{edward2024comprehensive}
J.~Edward, M.~M. Rosli, and A.~Seman, ``A comprehensive analysis of a framework for rebalancing imbalanced medical data using an ensemble-based classifier,'' \emph{Pertanika Journal of Science and Technology}, vol.~32, no.~6, 2024.

\bibitem{gutman2016skin}
D.~Gutman, N.~C. Codella, E.~Celebi, B.~Helba, M.~Marchetti, N.~Mishra, and A.~Halpern, ``Skin lesion analysis toward melanoma detection: A challenge at the international symposium on biomedical imaging (isbi) 2016, hosted by the international skin imaging collaboration (isic),'' \emph{arXiv preprint arXiv:1605.01397}, 2016.

\bibitem{sandler2018mobilenetv2}
M.~Sandler, A.~Howard, M.~Zhu, A.~Zhmoginov, and L.-C. Chen, ``Mobilenetv2: Inverted residuals and linear bottlenecks,'' in \emph{Proceedings of the IEEE conference on computer vision and pattern recognition}, 2018, pp. 4510--4520.

\bibitem{tomek1976two}
I.~Tomek, ``Two modifications of cnn.'' 1976.

\bibitem{mani2003knn}
I.~Mani and I.~Zhang, ``knn approach to unbalanced data distributions: a case study involving information extraction,'' in \emph{Proceedings of workshop on learning from imbalanced datasets}, vol. 126, no.~1.\hskip 1em plus 0.5em minus 0.4em\relax ICML United States, 2003, pp. 1--7.

\bibitem{yan2022spatial}
Y.~Yan, Y.~Zhu, R.~Liu, Y.~Zhang, Y.~Zhang, and L.~Zhang, ``Spatial distribution-based imbalanced undersampling,'' \emph{IEEE Transactions on Knowledge and Data Engineering}, vol.~35, no.~6, pp. 6376--6391, 2022.

\bibitem{chawla2002smote}
N.~V. Chawla, K.~W. Bowyer, L.~O. Hall, and W.~P. Kegelmeyer, ``Smote: synthetic minority over-sampling technique,'' \emph{Journal of artificial intelligence research}, vol.~16, pp. 321--357, 2002.

\bibitem{he2008adasyn}
H.~He, Y.~Bai, E.~A. Garcia, and S.~Li, ``Adasyn: Adaptive synthetic sampling approach for imbalanced learning,'' in \emph{2008 IEEE international joint conference on neural networks (IEEE world congress on computational intelligence)}.\hskip 1em plus 0.5em minus 0.4em\relax Ieee, 2008, pp. 1322--1328.

\bibitem{galar2011review}
M.~Galar, A.~Fernandez, E.~Barrenechea, H.~Bustince, and F.~Herrera, ``A review on ensembles for the class imbalance problem: bagging-, boosting-, and hybrid-based approaches,'' \emph{IEEE Transactions on Systems, Man, and Cybernetics, Part C (Applications and Reviews)}, vol.~42, no.~4, pp. 463--484, 2011.

\bibitem{10757164}
M.~Gun and G.~Bilgin, ``Classification of skin lesions using deep learning and machine learning methods,'' in \emph{2024 Innovations in Intelligent Systems and Applications Conference (ASYU)}, 2024, pp. 1--6.

\bibitem{al2025skin}
A.~Al~Shafi, A.~Muntakim, P.~C. Shill, R.~Zannat, and A.~Al-Amin, ``Skin lesion classification using a soft voting ensemble of convolutional neural networks,'' in \emph{2025 International Conference on Electrical, Computer and Communication Engineering (ECCE)}.\hskip 1em plus 0.5em minus 0.4em\relax IEEE, 2025, pp. 1--6.

\end{thebibliography}

\end{document}